# In-situ wavelength tuning of quantum-dot single-photon sources integrated on a CMOS silicon chip

Ryota Katsumi[1,2,a)], Yasutomo Ota[3,b)], Alto Osada[3], Takeyoshi Tajiri[1], Takuto Yamaguchi[1], Masahiro Kakuda[3], Satoshi Iwamoto[1,3], Hidefumi Akiyama[2] and Yasuhiko Arakawa[3]

[1] *Institute of Industrial Science, The University of Tokyo, 4-6-1 Komaba, Meguro-ku, Tokyo, Japan*

[2] *Institute for Solid State Physics, The University of Tokyo, 5-1-5 Kashiwanoha, Kashiwa, Chiba, Japan*

[3] *Institute for Nano Quantum Information Electronics, The University of Tokyo, 4-6-1 Komaba, Meguro-ku, Tokyo, Japan*

**Abstract**

**Silicon quantum photonics provides a promising pathway to realize large-scale quantum photonic integrated circuits (QPICs) by exploiting the power of complementary-metal-oxide-semiconductor (CMOS) technology. Toward scalable operation of such silicon-based QPICs, a straightforward approach is to integrate deterministic single-photon sources (SPSs). To this end, hybrid integration of deterministic solid-state SPSs, such as those based on InAs/GaAs quantum dots (QDs), is highly promising. However, the spectral and spatial randomness inherent in the QDs pose a serious challenge for scalable implementation of multiple identical SPSs on a silicon CMOS chip. To overcome this challenge, we have been investigating a new hybrid integration technique called transfer printing, which is based on a pick-and-place operation and allows for the integration of desired QD SPSs on any locations on the silicon CMOS chips at will. Nevertheless, even in this scenario, in-situ fine tuning for perfect wavelength matching among the integrated QD SPSs will be required for interfering photons from the dissimilar sources. Here, we demonstrate in-situ wavelength tuning of QD SPSs integrated on a CMOS silicon chip. To thermally tune the emission wavelengths of the integrated QDs, we augmented the QD SPSs with optically driven heating pads. The integration of all the necessary elements was performed using transfer printing, which largely simplified the fabrication of the three-dimensional stack of micro/nanophotonic structures. We further demonstrate in-situ wavelength matching between two dissimilar QD sources integrated on the same silicon chip. Our transfer-printing-based approach will open the possibility for realizing large-scale QPICs that leverage CMOS technology.**

---

a) E-mail: katsumi@iis.u-tokyo.ac.jp

b) E-mail: ota@iis.u-tokyo.ac.jp

Quantum photonic integrated circuits (QPICs) are of great research interest as a fascinating route toward the construction of future quantum photonic devices[1]. For the realization of large-scale QPICs, silicon quantum photonics, which exploits silicon photonics for quantum applications, is promising because of its direct access to large scale PICs using complementary-metal-oxide-semiconductor (CMOS) technology. The silicon-based PIC technology also offers well-developed nanophotonic elements[2–4] (e.g. high-Q micro-ring resonators, ultra-low-loss waveguides, optical beam splitters and modulators) that enable us to build highly functional and robust QPICs. There are several demonstrations of complex QPICs that have been empowered by matured CMOS technology[5–8].

Toward scalable operation of single-photon-based QPICs, it is of great importance to implement single-photon sources (SPSs) with the ability to deterministically emit indistinguishable single photons[9]. For this purpose, one possible approach is the hybrid integration of solid-state quantum emitters[10], such as carbon nanotube[11], color centers in diamond[12,13] and two-dimensional layered materials[14,15]. Among solid-state SPSs, InAs/GaAs self-assembled quantum dots (QDs) are especially promising because of their proven potential for the deterministic operation with pure and indistinguishable single-photon generation[16–19]. Furthermore, their emission wavelengths can be engineered to the telecom-wavelengths[20–23], where silicon is transparent. So far, encouraging progress has been made in the hybrid integration of QD-based SPSs onto silicon-based photonic platforms[24–27]. However, there remains a challenge in implementing desired QD SPSs on preferred locations on chip, which is hindered by position and spectral randomness inherent in the epitaxially-grown QDs. To overcome this challenge, we have been investigating a new hybrid integration approach based on transfer printing[28–31], which is a simple pick-and-place assembly technique using a transparent rubber stamp. By pre-characterizing SPSs prior to the transfer process, this approach may enable us to integrate multiple QD SPSs with desired optical properties on arbitrary locations on CMOS silicon photonic chips at will. Nevertheless, perfect spectral matching among integrated SPSs is still mandatory for the scalable operation of dominant quantum photonic information processing protocols, that rely on interferences among photons. Several methods have been studied to control emission wavelengths of QDs, such as thermal tuning[32], strain tuning[33,34], magnetic tuning[35] and Stark tuning[36,37]. We note that two-photon interference from independent QD SPSs on a single GaAs chip has recently been reported[38], but the interference was observed off-chip and the SPSs were not integrated into a photonic circuit.

In this work, we demonstrate in-situ fine wavelength tuning of QD SPSs integrated on a CMOS silicon chip. The emission wavelengths of the integrated QDs were thermally controlled using heating pads that absorb the heating laser irradiated from the above[32,38]. The heating pads were additionally integrated onto the silicon-integrated QD SPS. All the hybrid integration processes were performed using transfer printing[39–43], which facilitated the three-dimensional (3D) stack of the photonic micro/nano-structures. With this approach, we also demonstrate wavelength matching between two dissimilar QD sources integrated on the same silicon chip. Our transfer-printing-based approach will potentially enable the scalable integration of multiple identical QD SPSs on a CMOS silicon chip.

Figure 1 illustrates the structure of the investigated device. A QD SPS is placed above a glass-cladded silicon waveguide. A high-Q nanobeam cavity based on a GaAs-based one-dimensional (1D) photonic crystal (PhC) is utilized to support the efficient coupling of the QD emission into the silicon waveguide. To thermally tune the wavelength of the QD emission, a metal pad based on Cr is implemented on the marginal region of the integrated SPS. This pad can function as a heater when irradiated with an external heating laser beam, allowing for fine tuning of the wavelength of the QD emission on chip. The employed device structure was designed to achieve near-unity single-photon coupling of the QD into the silicon waveguide by tuning the vertical cavity-waveguide distance ($d$)[39,40,44]. For $d$ = 360 nm, the device supports a calculated emitter-waveguide coupling efficiency of 99.3%.

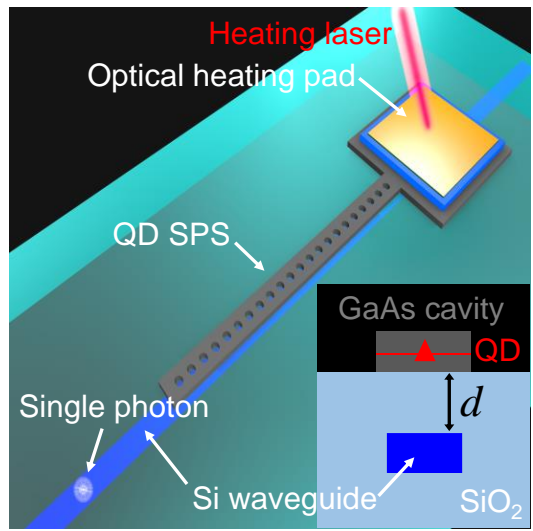

FIG. 1. Schematic of the investigated device structure. A QD SPS is placed above a glass-cladded silicon waveguide. An optical heating pad is implemented on the marginal region of the integrated SPS (Inset: schematic of the device cross-section).

The procedure of fabricating the device based on transfer printing is presented in Fig. 2. We started with the integration of an InAs/GaAs QD SPS onto a silicon waveguide [Figs. 2(a)-(d)]. We fabricated 1D PhC-based QD SPSs (hole radius = 78 nm, hole period = 300 nm, width = 450 nm, beam length = 13.5 µm) into a 180 nm-thick GaAs slab embedding one layer of self-assembled InAs QDs using the standard semiconductor nanofabrication processes (electron beam lithography and dry and wet etching processes) [Fig. 2(a)]. The hole period around the cavity center was modulated in order to support a high-Q factor[45–47]. In parallel, glass-cladded silicon waveguides were prepared using a CMOS foundry. The waveguide width is 250 nm with a thickness of 210 nm [Fig. 2(b)]. Under these parameters, the evanescent coupling between the cavity and the waveguide was maximized at the cavity resonant wavelength of around 1,170 nm[39]. The silicon waveguide was cladded by a 2 µm-thick $SiO_2$ layer through chemical vapor deposition during the foundry process. The thickness of the $SiO_2$ layer above the silicon waveguide (= $d$) was reduced to 360 nm using a dry etching process for efficient single photon waveguiding. The silicon waveguide is terminated by the exit ports, that allow the extraction of waveguide-coupled QD emissions into free space. We then integrated the prepared QD SPS onto the silicon waveguide by means of transfer printing. We picked a prepared QD SPS by attaching a polydimethylsiloxane (PDMS) transparent rubber stamp on it and then by rapidly pealing the stamp off [Fig. 2(c)]. The picked-up QD SPS was subsequently transferred to the silicon CMOS chip by placing the SPS on the top of the silicon waveguide and slowly releasing the transparent rubber stamp [Fig. 2(d)].

Next, we implemented an optical heating pad onto the integrated SPS. We prepared air-bridged plates (width = 7.8 µm, length = 24 µm) into a 220 nm-thick silicon slab using the aforementioned semiconductor nanofabrication processes [Fig. 2(e)]. Then we evaporated 20 nm-thick Cr on the silicon plates, which absorbs heating laser efficiently [Fig. 2(f)]. The separate fabrication of the heaters and SPSs allows us to maintain optical properties of the QD SPSs as the direct evaporation of Cr onto the QD SPS structure may degrade the optical quality of the QDs. The heating pad was transferred onto the marginal region of the integrated SPS by repeating the pick-and-place transfer printing process [Figs. 2(g) and (h)]. The insets in Fig. 2 show microscope images taken at each step of the process and Fig. 2(i) shows a microscope image of a completed device. All the assembling processes were performed using the homemade transfer-printing apparatus, that composed of precision motion-controlled piezo stages and an optical microscope[39]. In the current setup, both processes of picking-up and placement can be done with the success rates of close to 100%. The position deviation between the nanobeam and waveguide was deduced as < 100 nm, which is routinely possible with our transfer-printing system[39]. We emphasize that the simple pick-and-place operation of transfer printing largely facilitates the construction of the complicated 3D stack of micro/nanostructures.

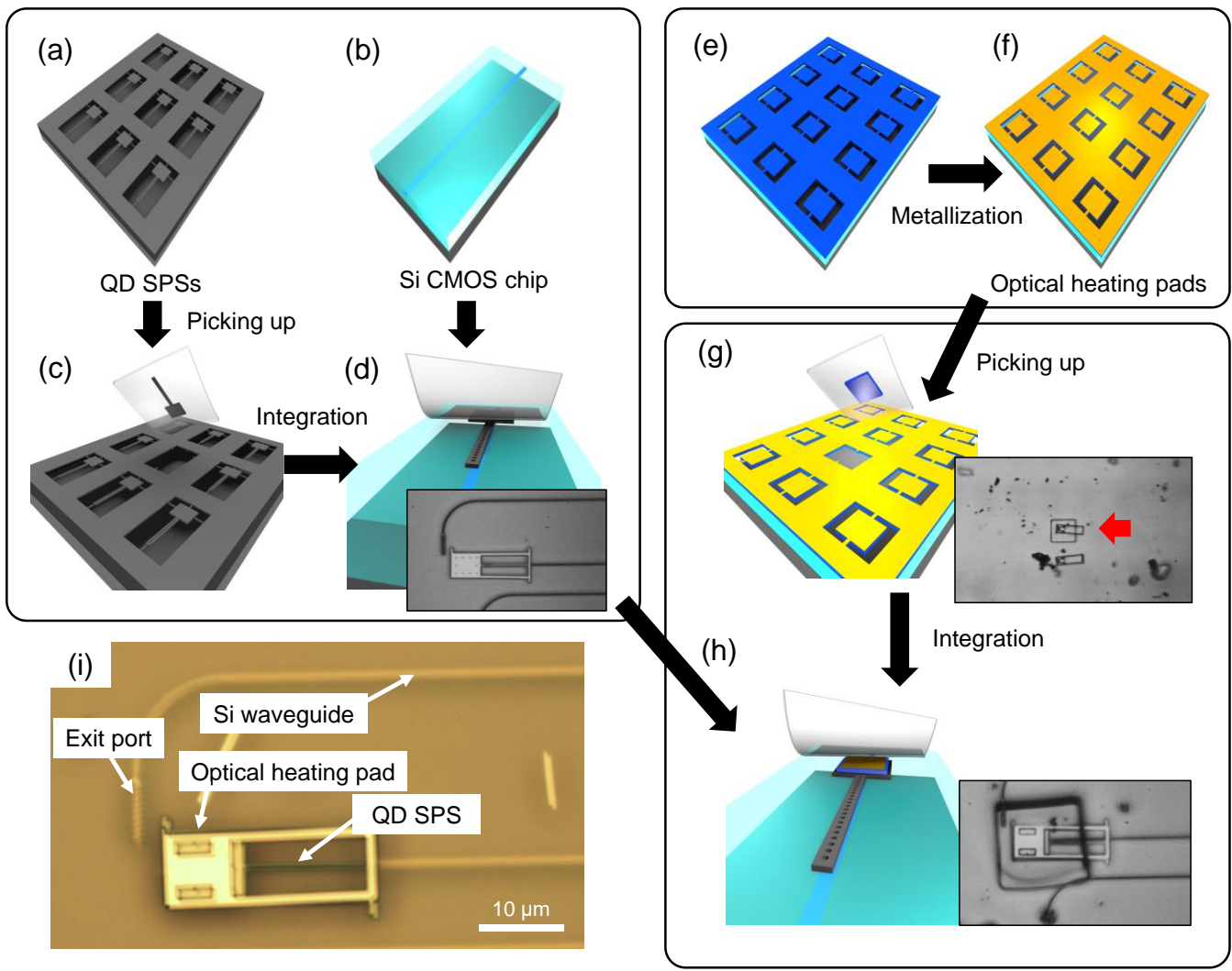

FIG. 2. Procedure of fabricating the device based on transfer printing. We started with the preparation of (a) InAs/GaAs QD SPSs based on PhC nanobeam cavities and (b) glass-cladded silicon waveguides. (c) Picking a prepared QD SPS by placing a PDMS transparent rubber stamp on it and then quickly pealing the transparent rubber stamp off. (d) The picked-up QD SPS was then accurately transferred to the top of the silicon waveguide by placing the SPS and slowly peeling off the transparent rubber stamp. We prepared (e) air-bridged silicon plates to fabricate (f) optical heating pads by evaporating 20-nm-thick Cr on the silicon plates. (g) and (h) The heating pad was then transferred onto the marginal region of the integrated SPS by repeating the pick-and-place transfer printing process. (i) Microscope image of a completed device. The insets show microscope images taken at each step of the process.

The fabricated devices were characterized by low-temperature micro-photoluminescence (μPL) measurements. An objective lens with x50 magnification and 0.65 numerical aperture was utilized to image the devices, focus pump laser on the device, and collect PL signals. The collected PL signals were analyzed using a grating spectrometer and an InGaAs camera. We radiated a continuous wave diode laser oscillating at 785 nm (average pump power of 1 μW) to the cavity center. A PL spectrum was measured from the exit port at 21 K, as shown in Fig. 3(a). We observed a strong peak of cavity mode emission at 1161.9 nm, together with that from a QD at 1162.9 nm. From the spectrum in Fig. 3(a), the experimental Q-factor of the cavity coupled to the waveguide is deduced to be $Q_{exp}$ = 2,500. Meanwhile, cavities decoupled from waveguides exhibit Q-factors of $Q_{ave}$ = 12,500, which was deduced from 10 nanobeam cavities that were placed on plane glass and completely decoupled from silicon waveguides. Given these values, the experimental cavity-waveguide coupling efficiency was estimated to be 80%[39], suggesting efficient waveguiding of the cavity mode and was comparable to the results of our previous work[40].

Then we tested the tunability of the fabricated devices whilst optically pumping the QD for light emission. We activated the heating pad by irradiating a heating laser (a continuous wave Ti:sapphire laser oscillating at 920 nm). This wavelength was chosen to suppress the excitation of the QDs by the heating laser. When increasing the power of the heating laser, QD emission was indeed redshifted as displayed in a series of PL spectra in Fig. 3(b). The dependence of the QD detuning on the heating laser power is shown in Fig. 3(c). The maximum tuning range of QD emission wavelength in the setup of this study was 0.9 nm and half of what was previously reported for an air-bridge photonic crystal nanocavity[32]. The narrower tuning range of the current work is probably due to the heat dissipation into the glass clad from the GaAs layer. We presume that heat transfer from the heating pad to the GaAs layer is efficient, even for transfer printed devices as reported in previous work[30]. To show the compatibility of this local tuning technique with single-photon generation, we also conducted an intensity correlation measurement based on a Hanbury Brown-Twiss setup equipped with two superconducting nanowire single-photon detectors (overall system time resolution of ~ 50 ps). The inset of the Fig. 3(b) shows a normalized second-order correlation function $g^{(2)}(t)$ measured for the QD peak under heating with a heating laser power of 6 mW. An anti-bunching with $g^{(2)}(0) = 0.43$ at the zero delay time demonstrates single-photon generation from QD even when using the local tuning approach. The non-zero $g^{(2)}(0)$ value was probably a result of the background cavity emission stemming from other off-resonant QDs inside the cavity.

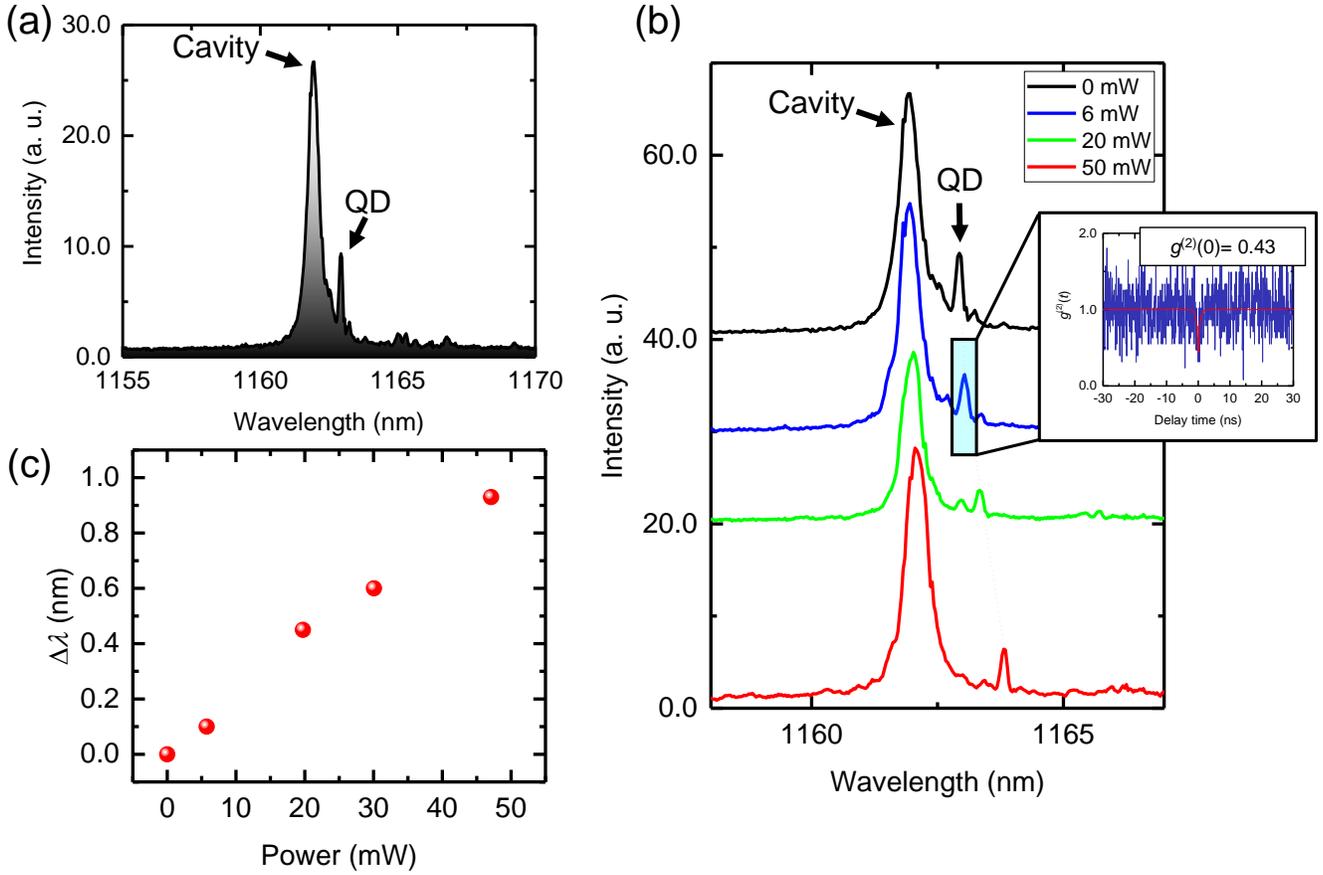

FIG. 3. (a) PL spectrum measured from the exit port at 21 K. (b) PL spectra measured from the exit port while changing the heating laser power (Inset: normalized second-order correlation function $g^{(2)}(t)$ measured for the QD peak under the spectral control). (c) Dependence of the QD detuning on the heating laser power.

Finally, we extended our method for the spectral matching between dissimilar integrated QDs, which is necessary for future demonstration of two photon interference on a silicon CMOS chip. For this purpose, we integrated two QD sources onto a single silicon circuit, together with optical heating pads by repeating the transfer printing process shown in Fig. 2. These waveguides are connected at a 50:50 beam splitter based on a 2 × 2 multimode interferometer waveguide (MMI) coupler. A microscope image of the completed sample is shown in Fig. 4(a). The two integrated QD sources were separately characterized using µPL measurements through the exit port. Strong QD emission peaks are observed for both sources as shown in Figs. 4(b) and 4(c) (which we label as QD1 and QD2, respectively). Then we simultaneously pumped both QD1 and QD2 and tuned QD2 with respect to QD1 using the heating pad for QD2. Figure 4(d) shows PL spectra collected at the exit port where both the signals from QD1 and QD2 exited at the same time. The emission from QD2 showed red shifts as the heating laser power increased. At a heating power of 28 mW, we observed wavelength matching between QD1 and QD2. Figure 4(e) summarizes the peak wavelengths of QD1 and QD2 as a function of heating laser power, which again confirms matching of the peak wavelength between the integrated two sources. This result suggests that mutual tuning of QD sources in proximity is possible using the current spectral tuning structure. The slight shift of QD1 was probably due to insufficient thermal separation from QD2, which can be circumvented by thermal insulation with a trench between the sources.

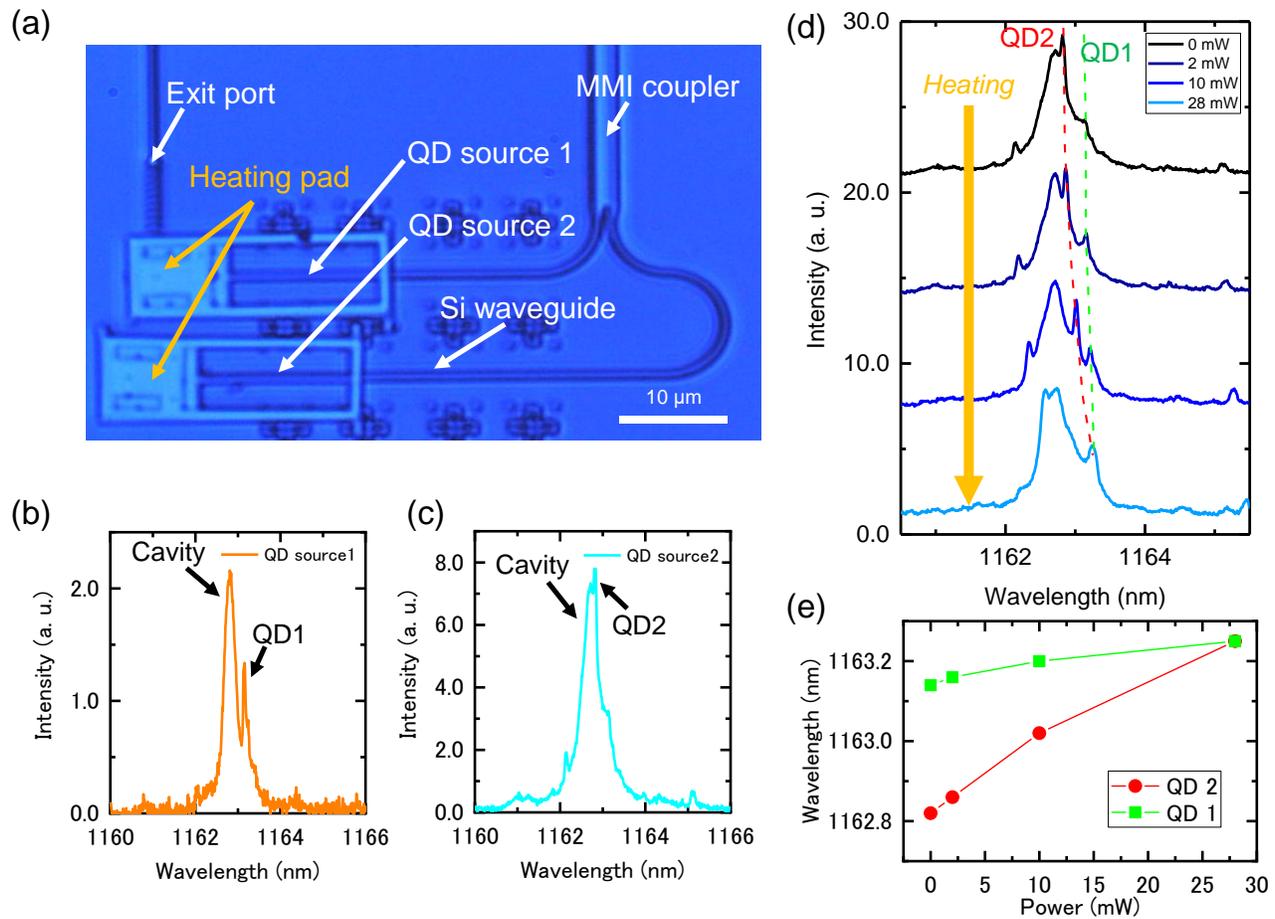

FIG. 4. (a) Microscope image of the completed sample. (b) and (c) PL spectra measured through the exit port. (d) PL spectra measured while irradiating the heating laser to the optical heating pad of QD source 2. The dotted red (green) lines show the peak positions of QD1 (QD2). (e) Emission wavelengths of QD1 and QD2 as a function of the heating laser power.

In summary, we have demonstrated in-situ wavelength tuning of integrated QD SPSs on a silicon CMOS chip. Optical heating pads were used to thermally control the emission wavelengths of the integrated QDs. All the required hybrid integration processes were conducted using transfer printing. We have further demonstrated the wavelength matching of two dissimilar integrated QD sources on the same silicon chip. These results suggest the potential for the scalable implementation of multiple identical QD SPSs onto a CMOS silicon photonic chip.

The authors thank S. Ishida and K. Kuruma for fruitful discussions. This work was supported by JSPS KAKENHI Grant-in-Aid for Specially Promoted Research (15H05700), KAKENHI (16K06294, 18J21667, 19K05300), JST PRESTO (JPMJPR1863), Inamori Foundation, Iketani Science and Technology Foundation, Murata Science Foundation and a project of the New Energy and Industrial Technology Development Organization (NEDO).